\documentclass[a4paper, 11points]{article}

\usepackage[a4paper, margin=2cm]{geometry}
\usepackage{url}
\usepackage{amsmath,amssymb}
\usepackage{bm, bbm}
\usepackage[FIGTOPCAP]{subfigure}
\usepackage{graphicx}
\usepackage{chngcntr}
\usepackage{comment}
\usepackage{natbib}

\usepackage{etoolbox}
\makeatletter
\patchcmd{\@makecaption}
  {\parbox}
  {\advance\@tempdima-\fontdimen2} 
  {}{}
\makeatother

\title{Spatial trend analysis of gridded temperature data at varying spatial scales}

\author{Ola Haug\footnote{ola.haug@nr.no}, Thordis L. Thorarinsdottir, Sigrunn H.  S{\o}rbye and Christian L. E. Franzke}

\begin{document}

\maketitle

\begin{abstract}
  \noindent
Classical assessments of trends in gridded temperature data perform
independent evaluations across the grid, thus, ignoring spatial correlations
in the trend estimates. In particular, this affects assessments of trend
significance as evaluation of the collective significance of individual tests
is commonly neglected. In this article we build a space-time hierarchical
Bayesian model for temperature anomalies where the trend coefficient is
modeled by a latent Gaussian random field. This enables us to calculate
simultaneous credible regions for joint significance assessments.  In a case
study, we assess summer season trends in 65 years of gridded temperature data
over Europe. We find that while spatial smoothing generally results in larger
regions where the null hypothesis of no trend is rejected, this is not the
case for all sub-regions.
\end{abstract}

\section{Introduction}

Analyses of temperature data in space and time play a crucial role in the study of climate change. Quantifying temperature trends globally and regionally is still part of the global warming debate and challenges policy makers and stakeholders in their efforts to postulate adequate adaptation and mitigation initiatives. In particular, sound and robust decision making calls for extending point estimates by also specifying their uncertainty \citep{thorarinsdottir:17}. Most assessments of changes in historical climate are performed by assessing trends in time series of observational data or observation-based data products, see e.g. the most recent assessment report of the Intergovernmental Panel on Climate Change (IPCC) \citep{AR5WG1}. The trend modeling is commonly performed by estimating the linear component of the change over time even if this simple approach has many well known shortcomings. Trends in the large-scale state of the climate system may
reflect systematic changes or low-frequency internal variability of the system, particularly for short time series \citep{vonStorchZwiers1999}. Longer time series may include breakpoints which change the nature of the trend \citep{SeidelLanzante2004}. Aiming for a more general approach to trend estimation, \citet{Wu&2007} define the trend as a monotonic function and suggest to estimate the trend by decomposing the data into so-called intrinsic mode functions in a non-parametric manner where the trend component is the residual (see also \citet{franzke2014warming}). Similarly, \citet{CraigmileGuttorp2011} build a wavelet-based hierarchical Bayesian model to estimate both seasonality and trend while \citet{Scinocca&2010} estimate a non-linear trend using smoothing splines.      

The trend assessment additionally requires an appropriate modeling of the
serial correlation in the data \citep{vonStorchZwiers1999,
  ChandlerScott2011}. In the most recent assessment of atmospheric and
surface observations, the IPCC chose to employ a linear trend model with a
first-order autoregressive, or AR(1), error structure following
\citet{Santer&2008} {\em ``because it can be applied consistently to all the
  data sets, is relatively simple, transparent and easily comprehended, and is
  frequently used in the published research assessed''}
\citep[pp. 180]{AR5WG1Ch2}. A comparison of various analysis methods for
global mean annual temperature series revealed that allowing AR(1) dependence
in the data yields confidence intervals for the trend estimates that are
roughly twice the width of those obtained under assumptions of independence
\citep{AR5WG1Ch2SM}. Alternative approaches to model the temporal correlation
include long-term memory models
\citep[e.g.][]{bunde2014significant,CraigmileGuttorp2011,franzke2010long,franzke2012nonlinear,ludescher2015long}
which are more appropriate for data with natural low-frequency variability \citep{lorenz1976nondeterministic}.    

A standard approach in climatology is to work with gridded data products
\citep[e.g.][]{new&2000}. For gridded data and multi-site analyses, the trend
modeling and the subsequent testing of trend significance is usually performed
independently in each grid point location \citep{AR5WG1Ch2}, with a few
exceptions \citep[e.g.][]{CraigmileGuttorp2011}. \citet{LivezeyChen1983} and
later \citet{wilks2006field, wilks2016stippling} rightfully argue that
collections of multiple statistical tests, such as individual tests at many
spatial grid points, are often interpreted incorrectly and in a way that
overstates research results. \citet{wilks2006field, wilks2016stippling}
suggests to control the false discovery rate \citep{benjamini1995controlling} to deal with this problem. We propose to take this a step further and perform a joint spatial analysis and construct simultaneous credible regions to assess the significance of the trend estimates jointly in space. 

We employ the same basic trend assessment model as the IPCC uses in its most
recent assessment report \citep{AR5WG1Ch2}. That is, we apply a Gaussian model
with a linear trend and an AR(1) temporal correlation structure. Additionally,
we assume a spatial structure in the trend coefficient and add a spatial error
term. Parameter estimation is obtained using Bayesian methodology, combining
the stochastic partial differential equation (SPDE) approach of
\citet{lindgren:11} with the methodology of integrated nested Laplace
approximation (INLA) \citep{rue:09}. Based on the posterior distribution of
the spatial trend coefficient, we then perform a simultaneous assessment over
the spatial domain following \citet{BolinLindgren2015}. We compare the joint
assessment to independent assessments in each grid point location based on the
marginal posterior distributions.  In a case study, we apply our approach to a
gridded data product for summer mean temperatures in Europe.  There is a
consensus that temperatures in Europe are generally warming. However, to which
extent depends on the region or spatial scale, the time frame and the data source considered \citep[e.g.][]{Bohm&2001, CraigmileGuttorp2011, franzke2015local, gao2017quantile, LorenzJacob2010, Tietavainen&2010, Schrier&2011, Schrier&2013}.

The remainder of this paper is organized as follows. The data used in our analysis is described in Section~\ref{sec:data}. The following Section~\ref{sec:methods} provides a description of the model, a review of the Bayesian estimation approach and a description of the methods for assessing the significance of the trend estimates. The results are provided in Section~\ref{sec:results} and conclusions are given in Section~\ref{sec:discussion}. 

\section{Data}\label{sec:data}

We analyze the E-OBS gridded daily temperature data product version 11.0 for the time period 1st January 1950 through 31st December 2014 \citep{Haylock&2008}. E-OBS is a land-only gridded version of the European Climate Assessment (ECA) data set which contains series of daily observations at meteorological stations throughout Europe and the Mediterranean. The original data product covers the area 25$^{\circ}$-75$^{\circ}$N $\times$ 40$^{\circ}$W-75$^{\circ}$E on a 0.25 degree regular grid. Prior to the statistical analysis, we compute a seasonal mean based on the daily data for the summer season covering June through August (JJA). This results in a time series of length 65 in each grid point. Finally, each time series is mean-centered and standardized prior to the analysis leaving series of local anomalies for investigation of trends on a decadal scale. To investigate the effects of spatial scale on the resulting trend estimates, we consider spatially up-scaled versions of the data where the 0.25 degree gridded anomalies have been up-scaled to a regular 1 degree grid and a regular 5 degree grid over the entire European domain as well as data on a regular 0.5 degree grid over Fennoscandia and Iberia.

\section{Methods}\label{sec:methods}
Let $\{Y_{{\bf s}t}\}$ denote a set of temperatures, in our case standardized seasonal mean temperature anomalies, at spatial locations ${\bf s} \in \{ {\bf s}_1, \ldots, {\bf s}_n \} \subset \mathbb{R}^2$ and time points $t \in \{ t_1, \ldots, t_m \} \subset \mathbb{R}_+$. The aim of the current study is to assess the spatial extent of potential trends in the data set. For this, we perform a spatial analysis in which both the model estimation and the subsequent significance testing are performed in a spatially coherent manner. 

\subsection{Linear trend estimation}
For a spatial linear trend estimation, we employ a space-time model of the type
\begin{align}\label{eq:obs}
  Y_{{\bf s}t} &=g_{\bf s}(t) + \varepsilon_{{\bf s}t}, \\
  \varepsilon_{{\bf s}t} &\sim N(0, \sigma_{\varepsilon}^2), \nonumber 
\end{align}
where $g_{\bf s}(t)$ describes the trend at location ${\bf s}$ while $\varepsilon_{{\bf s}t}$ denotes Gaussian noise which is uncorrelated in space and time. More explicitly, we assume the following specification for the trend term $g_{\bf s}(t)$,
\begin{align}\label{eq:spatial model}
g_{\bf s}(t) &= (\beta_0 + \beta_{\bf s})\,t + \tau_{{\bf s}t}, \\
\tau_{{\bf s}t} &= \varphi \tau_{{\bf s}(t-1)} + \xi_{{\bf s}t}, \nonumber \\
\bm{\beta} &\sim N(0,\bm{\Sigma}_\beta), \nonumber \\
\bm{\xi}_{t} &\sim N(0,\bm{\Sigma}_\xi), \nonumber
\end{align}
where $\bm{\beta} = (\beta_{{\bf s}_1}, \ldots, \beta_{{\bf s}_n})^\top$ and $\bm{\xi}_{t} = (\xi_{{\bf s}_1t}, \ldots, \xi_{{\bf s}_nt})^\top$ for $t \in \{t_1, \ldots, t_m\}$. That is, the trend $g_{\bf s}(t)$ follows an AR(1) process in time with noise terms that are  zero-mean and temporally independent but spatially correlated Gaussian random fields (GRFs). We denote the continuously indexed GRFs by $\{ \xi_t({\bf s}), \, {\bf s} \in \mathbb{R}^2\}_{t = t_1}^{t_m}$ where $\xi_t({\bf s}_i) = \xi_{{\bf s}_i t}$ for $i = 1, \ldots, n$. The model also assumes a spatially correlated GRF denoted $\{ \beta({\bf s}), \, {\bf s} \in \mathbb{R}^2\}$ for the trend coefficient, again with $\beta ({\bf s}_i) = \beta_{{\bf s}_i}$ for $i = 1, \ldots, n$. Specifically, we assume that each of the GRFs for the noise and trend coefficients has a stationary Mat\'ern covariance structure \citep{Matern1960}. This implies that the covariance between two components of the GRF at spatial locations ${\bf s}_i$ and ${\bf s}_j$ is modeled by the covariance function
\begin{equation}
c({\bf s}_i, {\bf s}_j) = \frac{\sigma^2}{2^{\nu - 1}\Gamma(\nu)} (\kappa d({\bf s}_i, {\bf s}_j))^{\nu} K_{\nu}(\kappa d({\bf s}_i,{\bf s}_j)), \label{eq:matern}
\end{equation}
where $\Gamma$ is the gamma function, $K_{\nu}$ is the modified Bessel function of the second kind, $d({\bf s}_i,{\bf s}_j)$ denotes the Euclidean distance between the locations ${\bf s}_i$ and ${\bf s}_j$, $\sigma^2$ is the marginal variance parameter, $\nu$ controls the smoothness and $\kappa$ is a spatial scale parameter.

\subsection{Statistical inference using the SPDE-INLA approach}
Spatial analyses involving high-dimensional covariance matrices are often not computationally feasible due to the dense structure of such matrices. A computationally efficient alternative in fitting \eqref{eq:obs}-\eqref{eq:spatial model} is to make use of the SPDE-approach introduced by \cite{lindgren:11}. A key idea of this approach is to construct continuously indexed approximations of GRFs by solving SPDEs. The solution is then represented as a  Gaussian Markov random field (GMRF) having a sparse precision (inverse covariance) matrix. The GMRF representation is used for practical computation and the solution retains a well-defined continuous interpretation. 

For illustration, we take a brief look at how the GMRF representation of the trend coefficient field is constructed. For general dimension $d$ let $\{\beta(\bm{s})$, $\bm{s}\in\mathbb{R}^d\}$ denote a continuously indexed GRF with Mat{\'e}rn covariance, also referred to as a Mat{\'e}rn field. This field represents an exact and stationary solution to the stochastic partial differential equation 
\begin{equation}
(\kappa^2-\nabla)^{\alpha/2} (\tau\beta(\bm{s}))={\cal W}(\bm{s}), \quad \bm{s}\in\mathbb{R}^d, \label{eq:spde}
\end{equation}
where  $\nabla$ is the Laplacian and ${\cal W}(\bm{s})$ is a Gaussian white noise process. The parameters in the two formulations \eqref{eq:matern} and \eqref{eq:spde} are coupled in that the Mat\'ern smoothness equals $\nu =\alpha - d/2$ and the marginal variance is given by
\begin{equation}\label{eq:variance}
\sigma^2 = \frac{\Gamma(\nu)}{\Gamma(\alpha) (4 \pi)^{d/2} \kappa^{2 \nu} \tau^2}.
\end{equation}
Furthermore, the range of the covariance structure can be described by
\begin{equation}\label{eq:range}
  \rho = (8 \nu)^{1/2} / \kappa,
\end{equation}
  the distance for which the correlation function has fallen to approximately 0.13, for all $\nu > 1/2$ \citep{LindgrenRue2015}. Here, we set $\alpha = 2$ which is the most natural choice for $d=2$ according to \cite{Whittle1954}. Alternative models are discussed in e.g. \citet{lindgren:11} and \citet{LindgrenRue2015}. 

\begin{figure}[!hbpt]
  \centering
  \subfigure[]{\includegraphics[height=6cm]{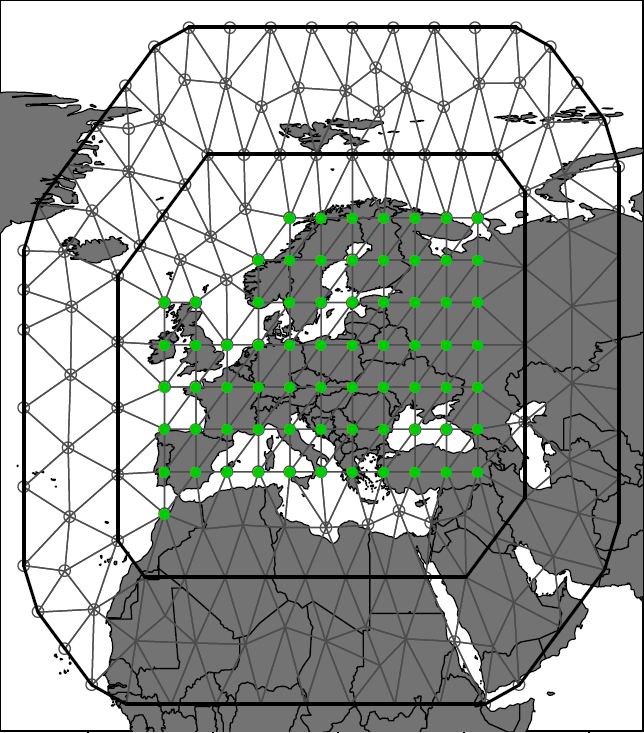}}
  \subfigure[]{\includegraphics[height=6cm]{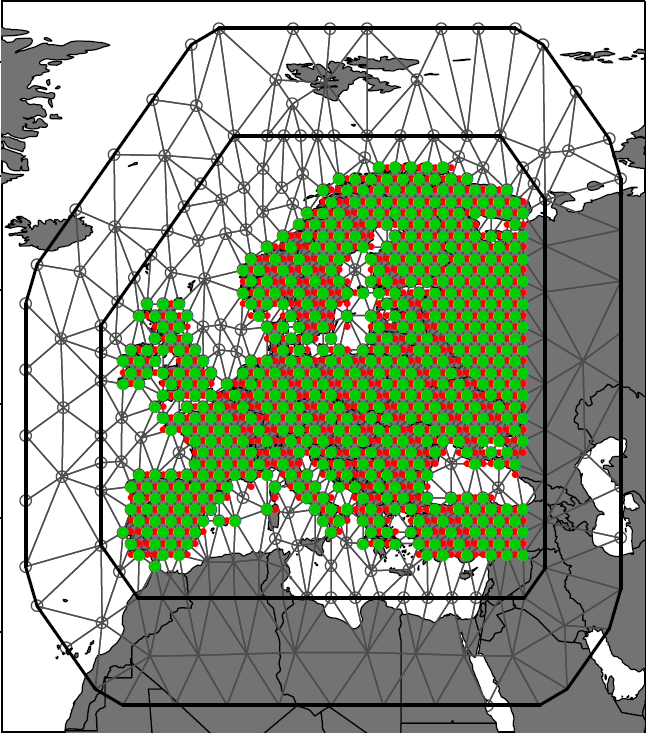}}
  \caption{The triangular meshes used in our full European domain analyses for data on a 5 degree grid (a) and data on a 1 degree grid (b). Data locations are indicated by red dots; those coinciding with mesh vertices superimposed in green.}\label{fig:mesh}
\end{figure}

An approximate solution of \eqref{eq:spde} can be found using the finite element method by first defining a triangular mesh with $G$ vertices on the relevant continuous domain and then use the piecewise linear representation
$$\beta(\bm{s})=\sum_{g=1}^G \psi_g(\bm{s}) \tilde{\beta}_g.$$
Here, the basis functions $\{\psi_g(\bm{s})\}$ are chosen to have the value 1 at vertex $g$ while being 0 at all other vertices. The  weights $\{\tilde{\beta}_g\}$, giving the heights of the triangles, have a Gaussian distribution with zero-mean. The resulting discretely-indexed vector for the trend coefficients at the vertices, 
$\tilde{\bm{\beta}}=(\tilde{\beta}_1,\ldots , \tilde{\beta}_G)$, will be a GMRF as the SPDE-solutions of \eqref{eq:spde} are Markovian when $\alpha$ is integer-valued,  see \cite{lindgren:11} for further details. Examples of the triangular mesh used in our analysis are shown in Figure~\ref{fig:mesh} and a list of the data sets used for the analysis is given in Table~1.

\begin{table}
  \caption{Overview over the data resolution, the size of the spatial data grid and the mesh size for the four data sets analyzed in this study.}
  \label{table:data}
  \centering
  \fbox{%
  \begin{tabular}{llrr}
    \hline
    Area & Resolution & Grid size & Mesh size \\
    \hline
    Europe & 5$^{\circ}$ & 70 & 203 \\
    Europe & 1$^{\circ}$ & 1213 & 711 \\
    Fennoscandia & 0.5$^{\circ}$ & 949 & 1153 \\
    Iberia & 0.5$^{\circ}$ & 324 & 430 \\
    \hline
    \end{tabular}}
\end{table}

A major benefit of the SPDE-approach is its implementation within the computational framework of latent Gaussian models using the \texttt{R}-package \texttt{R-INLA} \citep{rue:09, rue:17}. In general, latent Gaussian models represent a subclass of structured additive regression models \citep{fahrmeir:01} having a  three-stage hierarchical structure. First, the observations $\{Y_{{\bf s} t}\}$  are assumed conditionally independent given a latent field $\bm{x}$ and hyper-parameters $\bm{\theta}_1$. In our case, the likelihood is then specified by
$$\pi(\bm{y}\mid \bm{x},\bm{\theta}_1)=\prod_{j=1}^m\prod_{i=1}^n \pi(y_{{\bf s}_i t_j}\mid \bm{x},\bm{\theta}_1)$$
in which we assume a Gaussian distribution for the observations.  Second, all random variables in \eqref{eq:spatial model}, including the predictor $\eta_{{\bf s}t}=E(Y_{{\bf s}t})$,  are incorporated in the latent field $\bm x=\{\beta_0,\bm{\beta}, \{\bm{\xi}_t\}, \{\eta_{{\bf s}t}\}\}$.  A crucial assumption of latent Gaussian models is that $\bm{x}$ is a Gaussian Markov random field, i.e. 
$$\pi(\bm{x}\mid \bm{\theta}_2)\sim N(\bm{0},\bm{Q}^{-1}(\bm{\theta}_2)),$$ 
where the precision matrix $\bm{Q}(\bm{\theta}_2)$ is sparse. 
Finally, the third stage specifies a prior $\pi(\bm{\theta})=\pi(\bm{\theta}_1,\bm{\theta}_2)$ for the hyper-parameters, typically being non-Gaussian. In \eqref{eq:spatial model}, we have five hyper-parameters including the first-lag autocorrelation coefficient $\varphi$ and  the parameters $\tau$ and $\kappa$ related to the variance \eqref{eq:variance} and range \eqref{eq:range} of the Mat\'ern fields. The resulting posterior is then expressed by 
$$\pi(\bm{x},\bm{\theta}\mid\bm{y}) = \prod_{j=1}^m\prod_{i=1}^n \pi(y_{{\bf s}_i t_j}\mid \bm{x},\bm{\theta})\pi(\bm{x}\mid \bm{\theta})\pi(\bm{\theta}),$$
where the hyper-parameters are given independent priors.

The INLA-methodology is a deterministic approach that combines numerical approximations, interpolation and integration to provide accurate estimates of the posterior marginals for all components of the latent field and all the hyper-parameters. Implementation is greatly facilitated by the \texttt{R-INLA} package, which in general can be used to fit SPDE-models of different complexity and combine these with various latent model components to build the linear predictor, see \cite{cameletti2015} and \cite{krainski2018} for recent updates of the SPDE-INLA approach. 

\subsection{Assessing significance of trend estimates}

Denote by $\Omega \subset \mathbbm{R}^2$ our full study region, that is the union of the $n$ grid cells for which we have data. The uncertainty in the latent trend coefficient process $\bm{\beta}$ is commonly assessed using a credible band or an associated $p$-value under a null hypothesis of no trend, separately for each location ${\bf s} \in \Omega$ \citep{AR5WG1Ch2}. Such a pointwise credible band for $\bm{\beta}$ can be defined by the equi-tailed intervals $\{[q_{\alpha/2}({\bf s}_i), q_{1-\alpha/2}({\bf s}_i)]\}_{i=1}^{n}$, where $q_{\alpha}({\bf s}_i)$ denotes the $\alpha$-quantile of the posterior marginal distribution of $\beta_{{\bf s}_i}$. In our application, it is then of interest to assess for which locations ${\bf s} \in \{ {\bf s}_1, \ldots , {\bf s}_n\}$ the pointwise credible band does not contain the level $u = 0$. 

However, as such pointwise credible bands do not provide a joint interpretation, we also construct a simultaneous credible region so that with probability $1 - \alpha$, the trend coefficient field $\bm{\beta}$ stays inside the credible band at all spatial locations within the region. The simultaneous credible band is defined as the region $\{({\bf s},\beta)\, : \, {\bf s} \in \Omega, \, q_\rho({\bf s}) \leq \beta \leq q_{1-\rho}({\bf s}) \}$. Here, $\rho$ is chosen such that the posterior probability $\mathbbm{P}(q_\rho({\bf s}) < \beta_{\bf s}< q_{1-\rho}({\bf s}), {\bf s} \in \Omega) = 1 - \alpha$. Thus $\alpha$ controls the probability that the trend coefficient field is inside the credible band at all locations in $\Omega$. Similarly as above, we are here interested in the spatial region where  $u=0$ is not contained in the credible band. This region is also called the avoidance excursion set for the level $u=0$. The simultaneous credible bands and the associated avoidance excursion sets are calculated using the sequential integration method of \citet{BolinLindgren2015} as implemented in the {\tt R} package {\tt excursions}, see also \citet{BolinLindgren2016}.   

\section{Results}\label{sec:results}

Here, we present and compare the results for the four different data sets described in Table~\ref{table:data}. The code and data needed to reproduce the results for the 5 degree data set over Europe are available at \url{https://github.com/eSACP/STT}. 

\subsection{Fennoscandia and Iberia at 0.5 degree resolution}

\begin{figure}[!hbpt]
  \centering
  \subfigure[\hspace{10mm}]{\includegraphics[height=5cm]{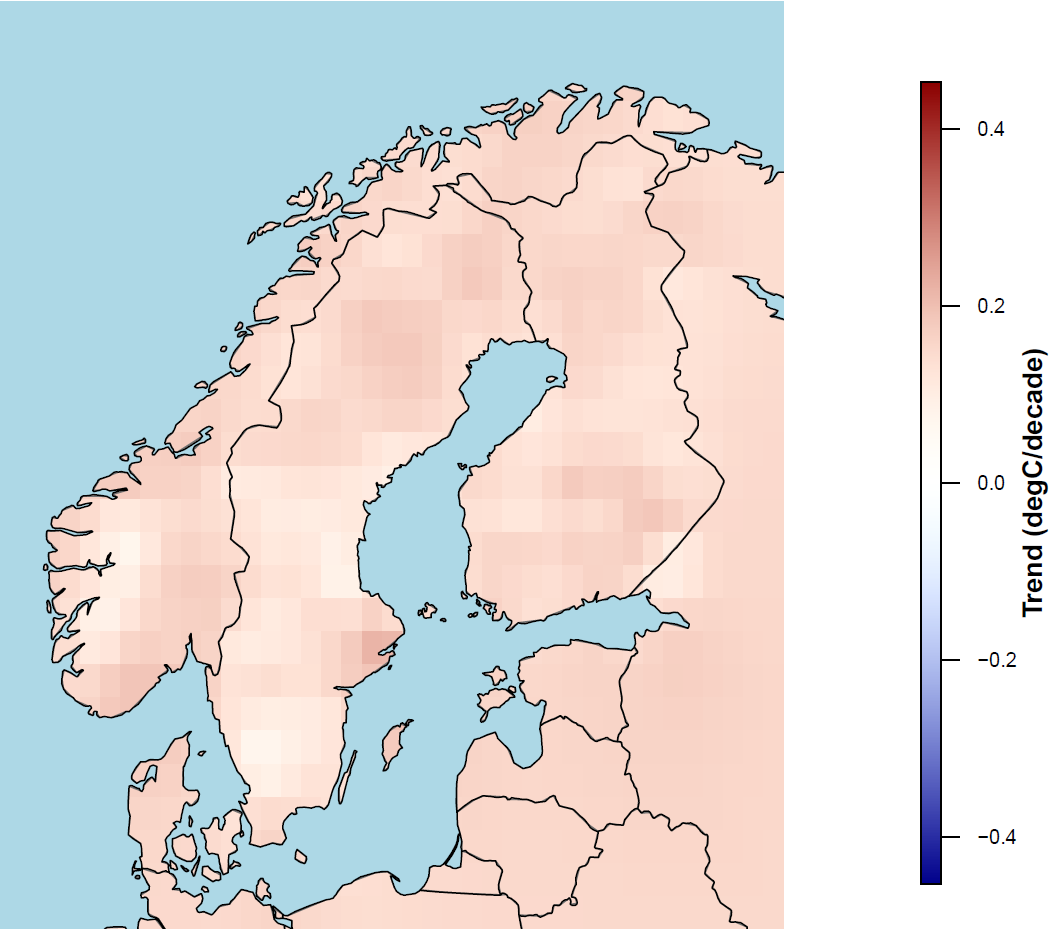}}
  \subfigure[\hspace{10mm}]{\includegraphics[height=5cm]{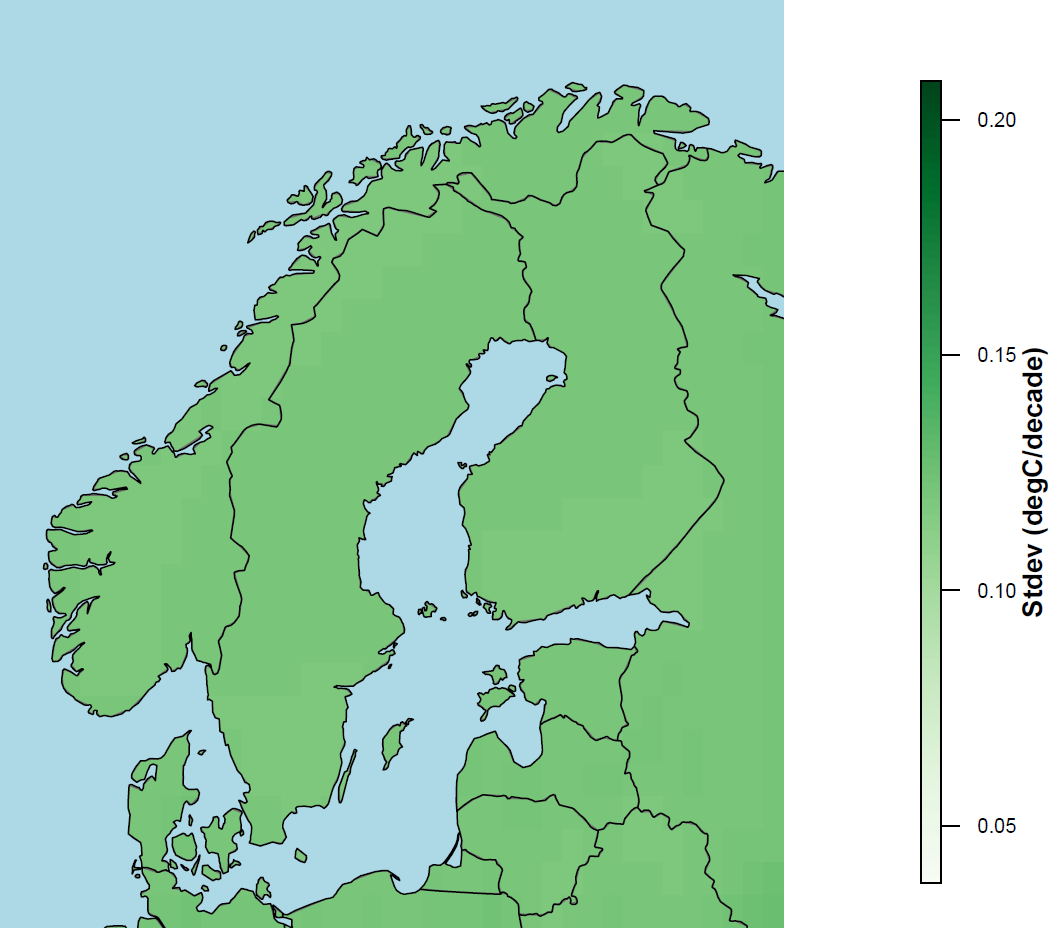}}
    \subfigure[]{\includegraphics[height=5cm, width=4.1cm]{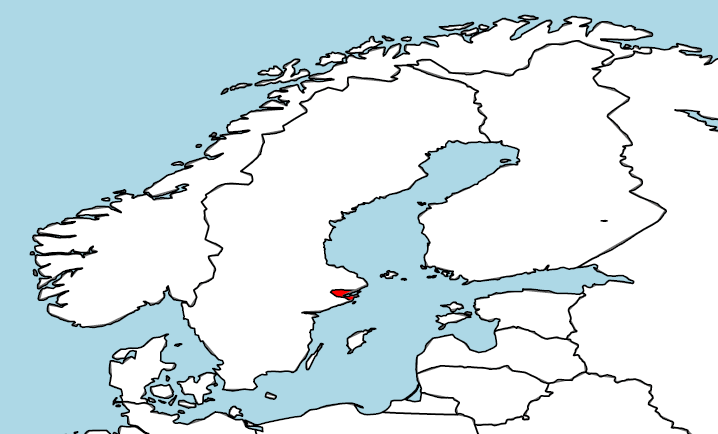}}
  \caption{Trend estimates for Fennoscandia based on data on a 0.5 degree grid: (a) Posterior mean estimates of the trend coefficient (in $^\circ$C/decade) projected on a regular 1 degree lattice; (b) Posterior standard deviation of the trend coefficient on the same lattice; (c) Avoidance excursion set where the spatial null hypothesis of no trend is rejected for level $\alpha = 0.05$ (red). Note that the estimates outside Fennoscandia are based on extrapolation of the data for these three countries.}\label{fig:Fenno}
\end{figure}

Figure~\ref{fig:Fenno} shows the results for data covering Fennoscandia, that is Norway, Sweden and Finland, on a 0.5 degree grid. The posterior mean of the trend estimates are positive overall and range from 0.08$^{\circ}$ to 0.21$^{\circ}$C per decade with the highest posterior mean close to Stockholm on Sweden's east coast. The posterior standard deviation is quite constant over the region at around 0.12$^{\circ}$C. As a result, the avoidance excursion set where the spatial null hypothesis of no trend is rejected at level $\alpha=0.05$ consists of a small region around Stockholm only. 

\begin{figure}[!hbpt]
  \centering
  \subfigure[\hspace{10mm}]{\includegraphics[height=5cm]{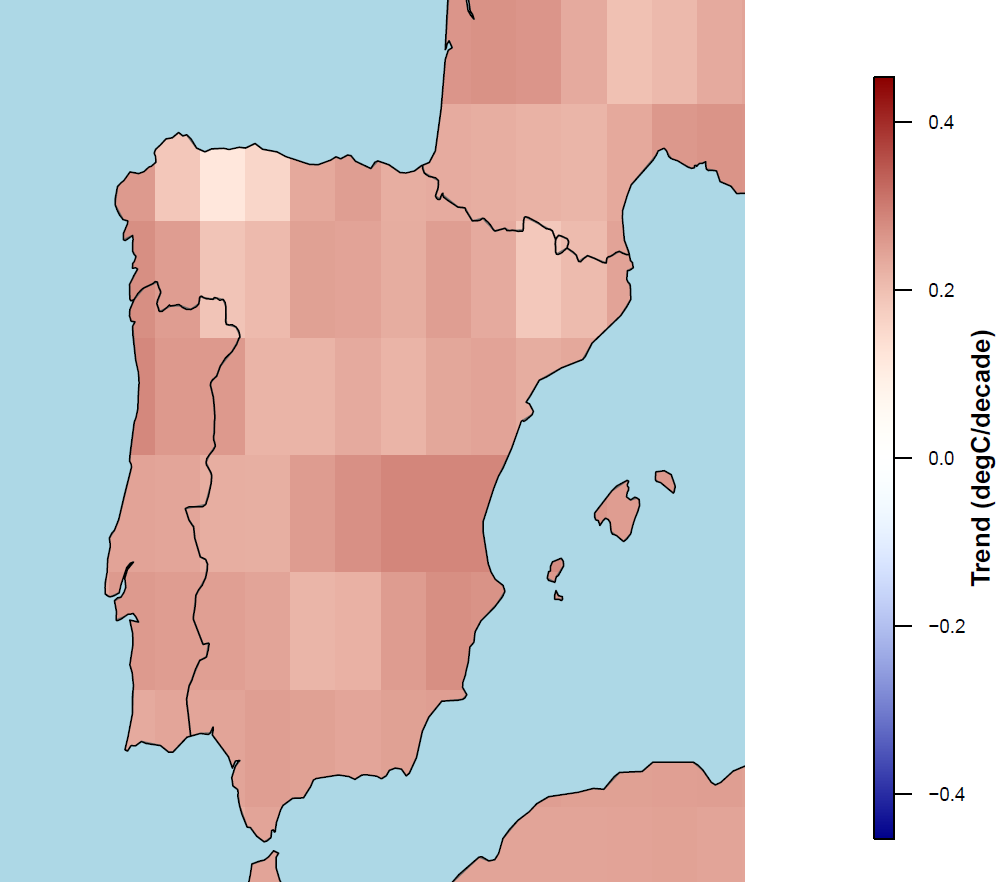}}
  \subfigure[\hspace{10mm}]{\includegraphics[height=5cm]{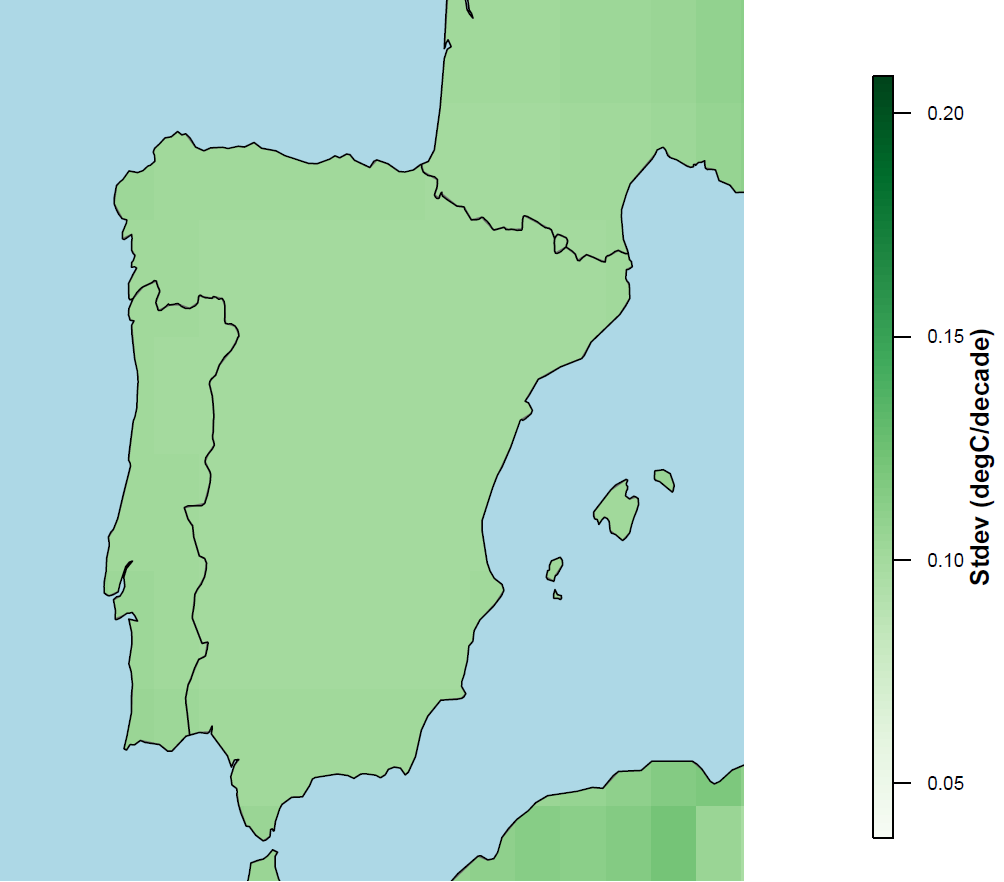}}
    \subfigure[]{\includegraphics[height=5cm, width=4cm]{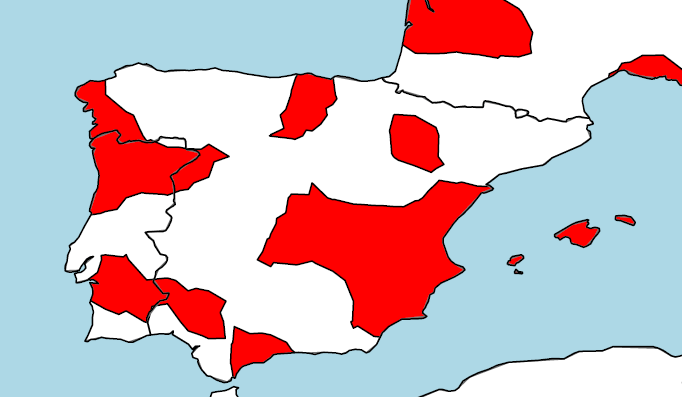}}
  \caption{Trend estimates for the Iberian Peninsula based on data on a 0.5 degree grid: (a) Posterior mean estimates of the trend coefficient (in $^\circ$C/decade) projected on a regular 1 degree lattice; (b) Posterior standard deviation of the trend coefficient on the same lattice; (c) Avoidance excursion set where the spatial null hypothesis of no trend is rejected for level $\alpha = 0.05$ (red). Note that the estimates outside Iberia are based on data extrapolation.}\label{fig:Iberia}
\end{figure}

The posterior mean trend estimates for the Iberian Peninsula in Figure~\ref{fig:Iberia} are somewhat higher than in Fennoscandia and range from 0.12$^{\circ}$ to 0.29$^{\circ}$C per decade. The lowest values occur in northern Spain while the highest trends are estimated in northern Portugal and in the Jucar river basin in eastern Spain. The posterior standard deviation of the trend estimates is somewhat lower than in Fennoscandia at approximately 0.10$^{\circ}$C per decade over the region. The avoidance excursion set for level $\alpha=0.05$ now covers slightly less than 50\% of the area indicating a greater degree of warming compared to Fennoscandia. Note that results for areas outside of the Iberian Peninsula are based on extrapolation of the data from Iberia and should thus be interpreted with care.  

\subsection{Europe at 5 and 1 degree resolutions}

\begin{figure}[!hbpt]
  \centering
  \subfigure[$5^\circ$ data grid]{\includegraphics[height=4.8cm]{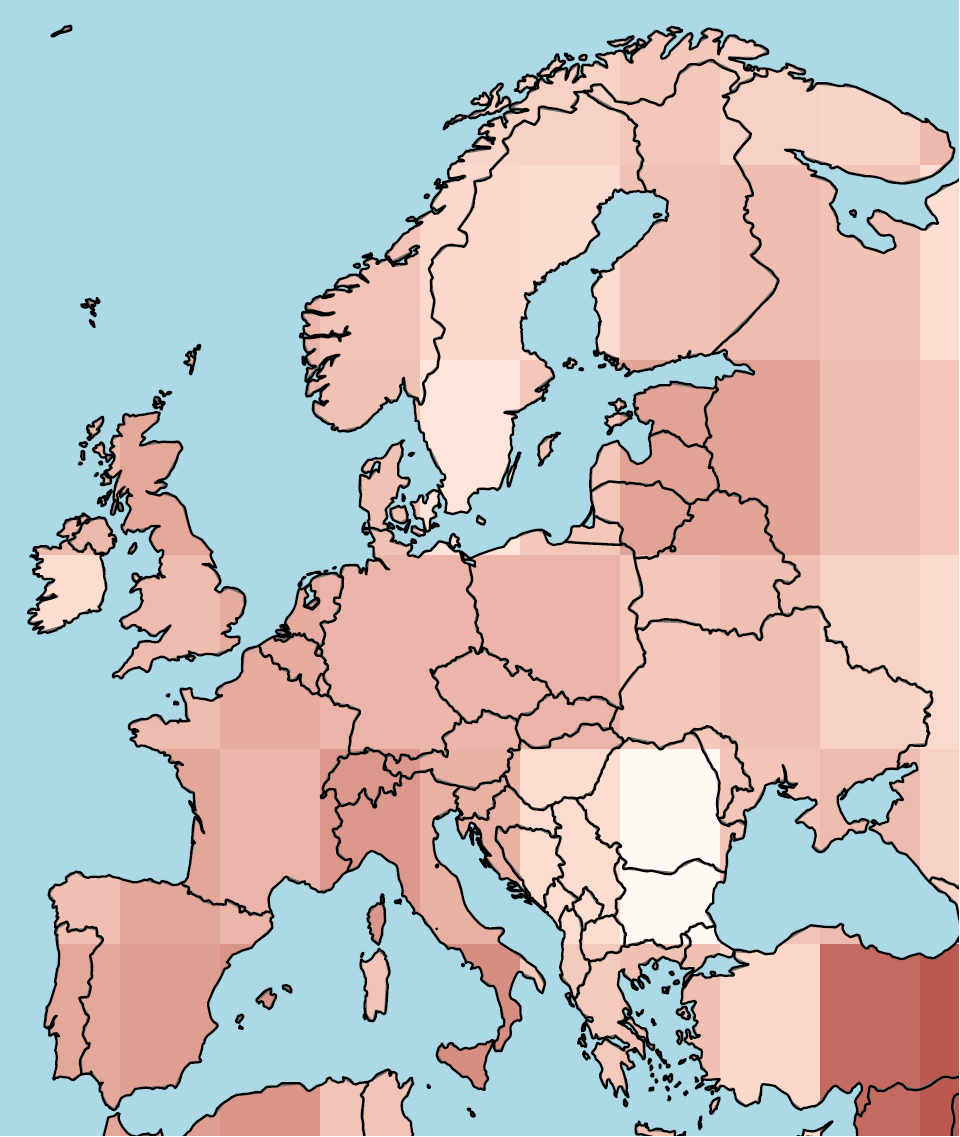}}
  \subfigure[$5^\circ$ data grid]{\includegraphics[height=4.8cm]{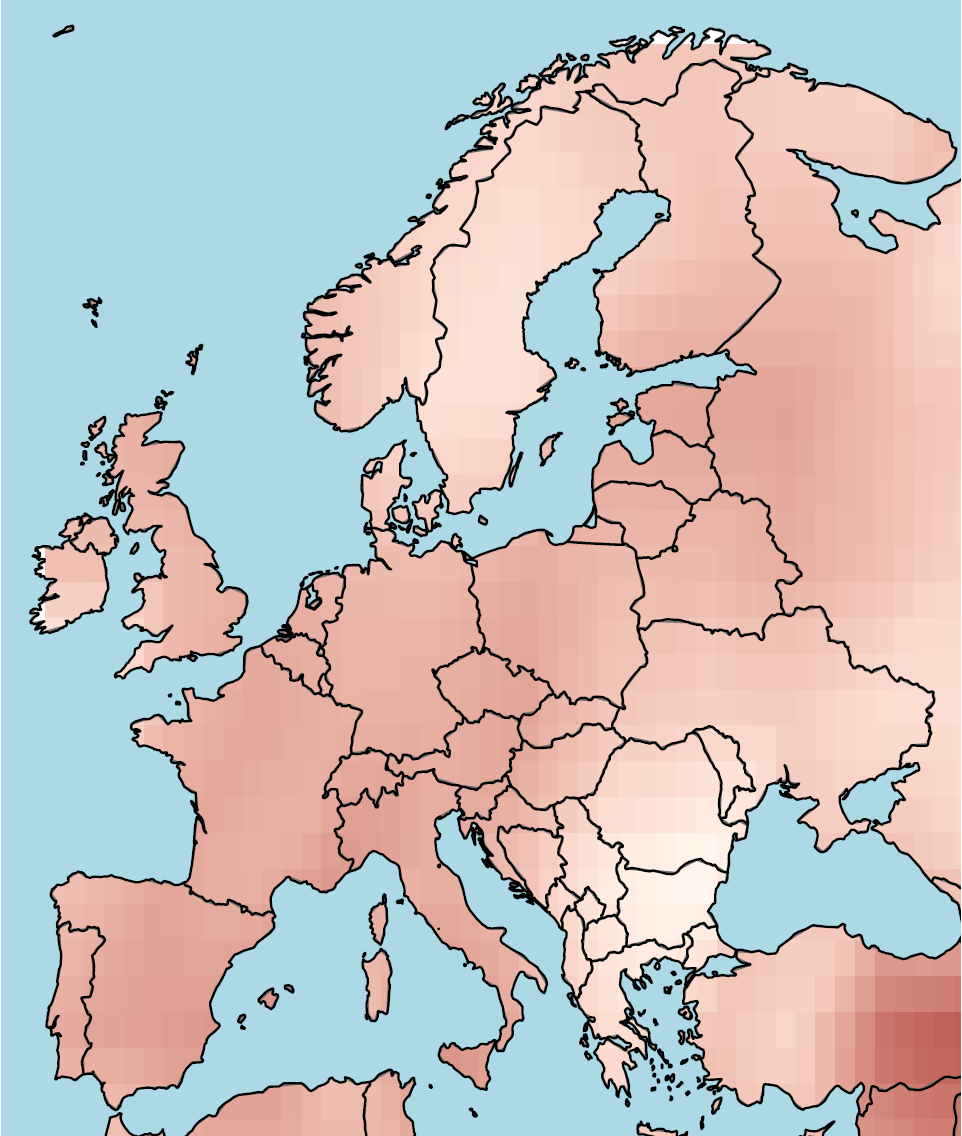}}
  \subfigure[$1^\circ$ data grid]{\includegraphics[height=4.8cm]{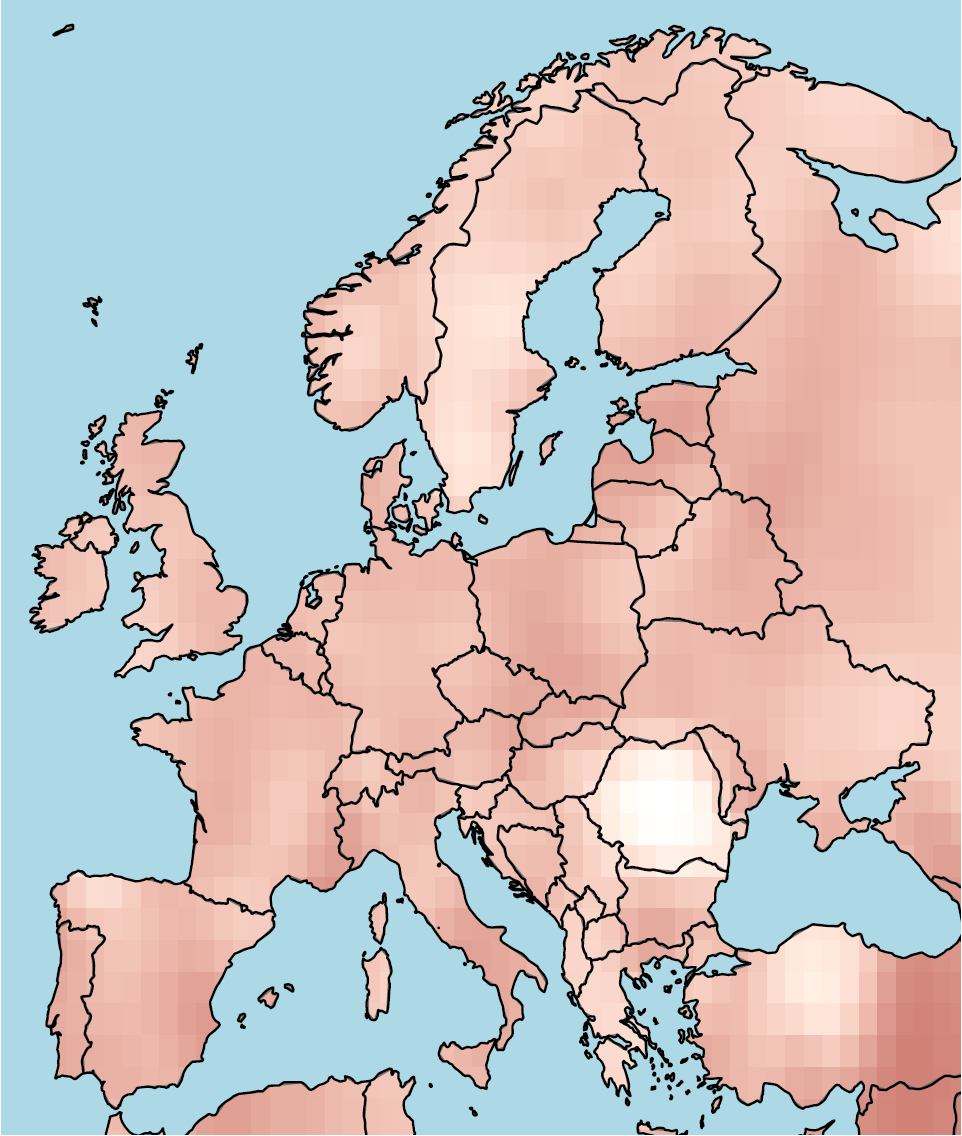}}
  \subfigure[$1^\circ$ data grid]{\includegraphics[height=4.8cm]{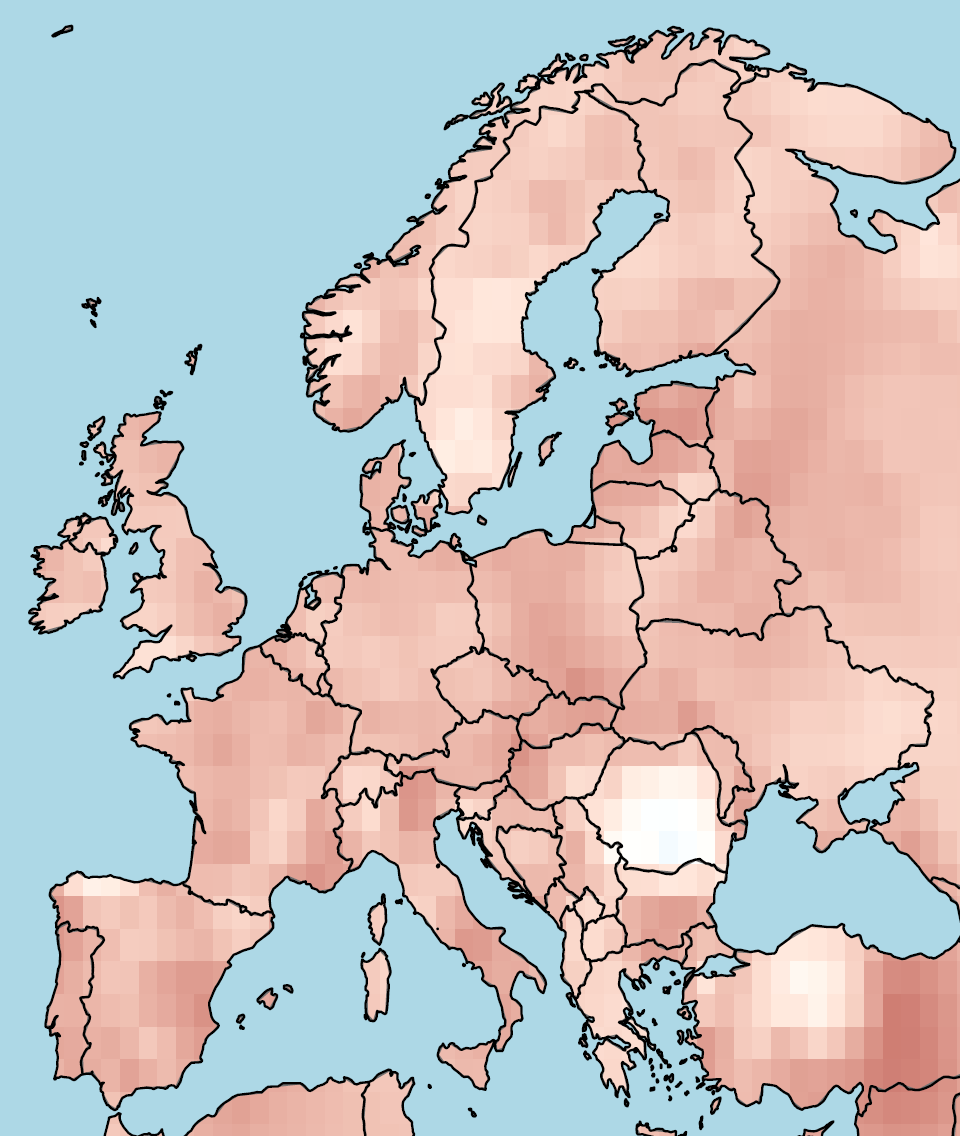}}
  \includegraphics[width=0.7\textwidth]{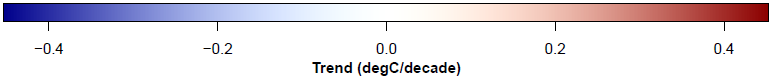}
    \caption{Posterior mean estimates of trends ($^\circ$C/decade) in the summer season 1950-2014 for 5 degree and 1 degree data grids resolutions. (a) and (c): Results on the respective meshes; (b) and (d): Results extrapolated to a 1 degree lattice.}\label{fig:trendEurope}
\end{figure}

Posterior mean estimates of trend coefficient fields for the coarser data grid resolutions of 5 and 1 degree are shown in Figure~\ref{fig:trendEurope}. For comparison, we consider results both on the original meshes used for the parameter estimation as well as extrapolated to a regular 1 degree lattice. For the 5 degree data, the posterior mean trend estimates on the lattice range between 0.07$^{\circ}$ and 0.34$^{\circ}$C per decade while the range is slightly lower at [-0.05, \, 0.30] for the higher 1 degree data resolution. At the coarser data resolution the lowest trends are estimated in Romania and Bulgaria in Eastern-Europe. When these estimates are extrapolated to the finer scale lattice, the lowest estimates seem mostly concentrated over Bulgaria. However, for a data resolution of 1 degree, the lowest trend estimates concentrate somewhat further north over Romania. Furthermore, we see that the finer resolution data results in a higher spatial variability in the trend estimates, as one might expect.   

Several other studies have assessed temperature trends in Europe using a range of methodologies and data sets. \cite{Tietavainen&2010} estimate a summer season warming of 0.13$^{\circ}$C per decade for 1959-2008 in Finland using spatially interpolated station data with a quadratic trend model with a 95\% confidence interval of (-0.08, \, 0.34). For comparison, our posterior mean estimates across lattice grid cells that cover Finland have a range of (0.19, \, 0.21) for the coarser 5 degree data resolution and a range of (0.17, \, 0.22) for the finer resolution. \cite{Schrier&2011} estimate a summer warming trend of 0.24$^{\circ}$C per decade for 1950-2008 in the Netherlands applying a linear trend model to the Central Netherlands Temperature time series. These estimates are slightly higher than those obtained here which for the 1 degree lattice cells covering the Netherlands have a range of (0.22, \, 0.23) when based on 5 degree data and a range of (0.17, \, 0.21) for 1 degree data. Some of these differences might be due to slightly different time periods considered. 

\begin{figure}[!hbpt]
  \centering
  \subfigure[$\alpha = 0.05$]{\includegraphics[height=4.8cm, width=4cm]{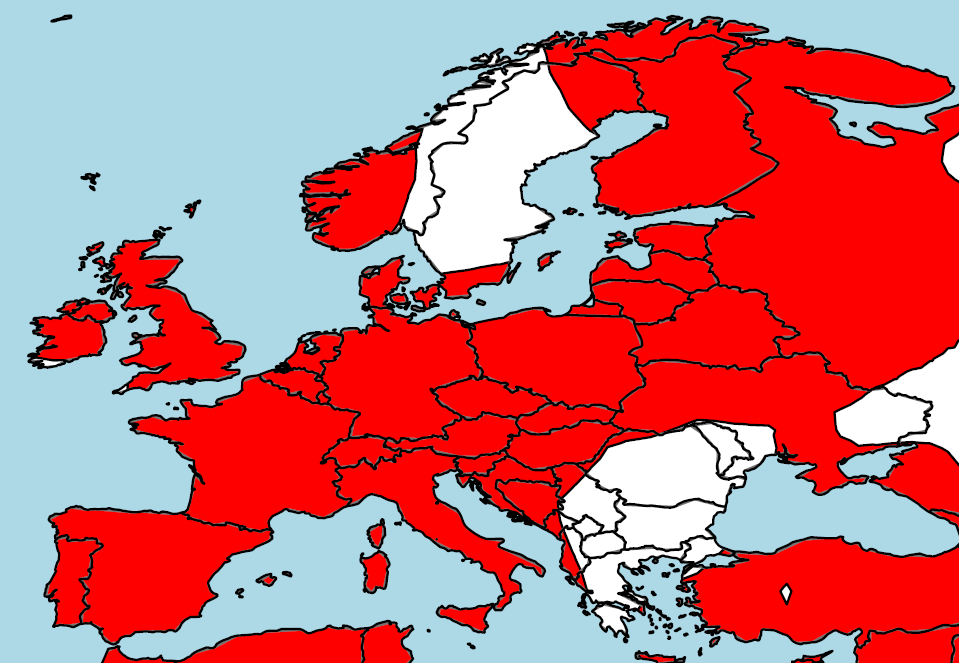}}
  \subfigure[$\alpha = 0.05$]{\includegraphics[height=4.8cm, width=4cm]{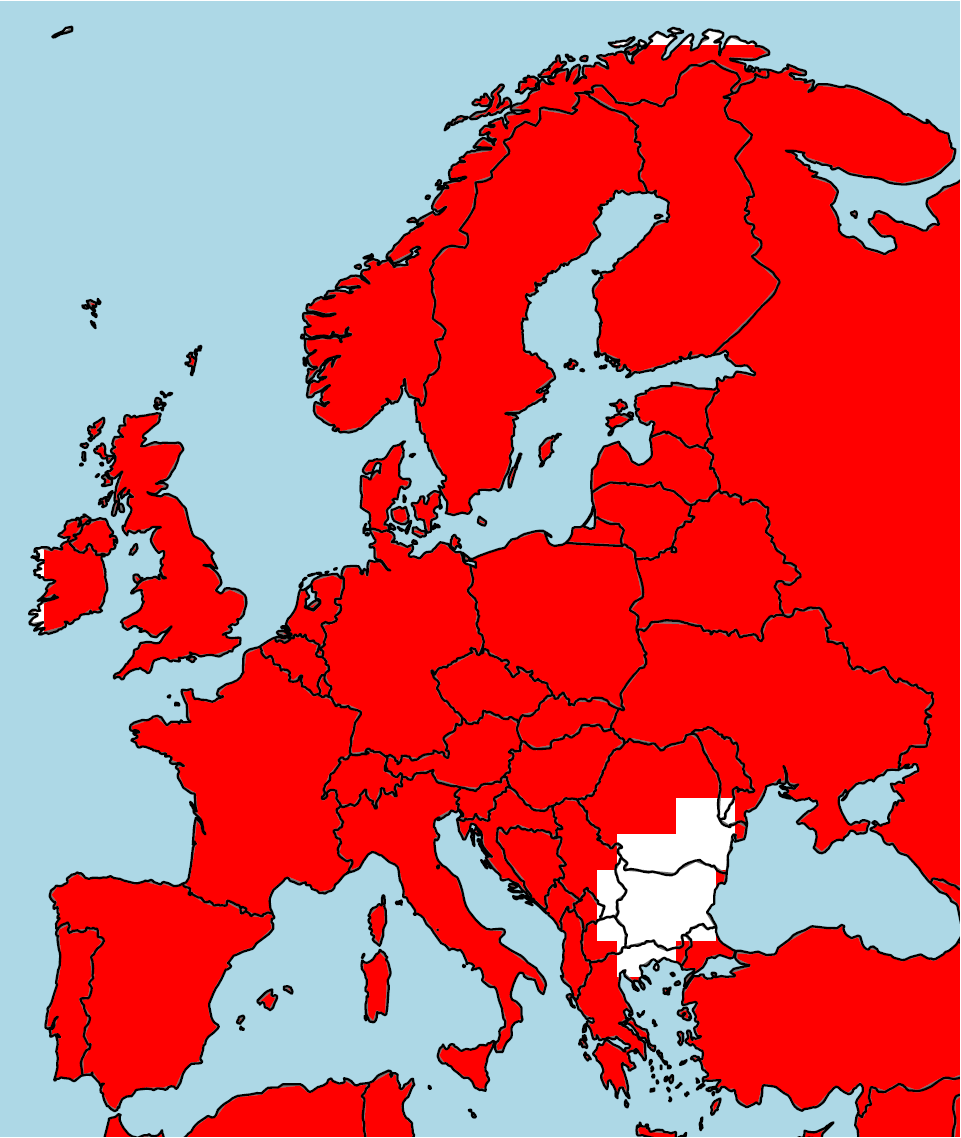}}
  \subfigure[$\alpha = 0.05$]{\includegraphics[height=4.8cm, width=4cm]{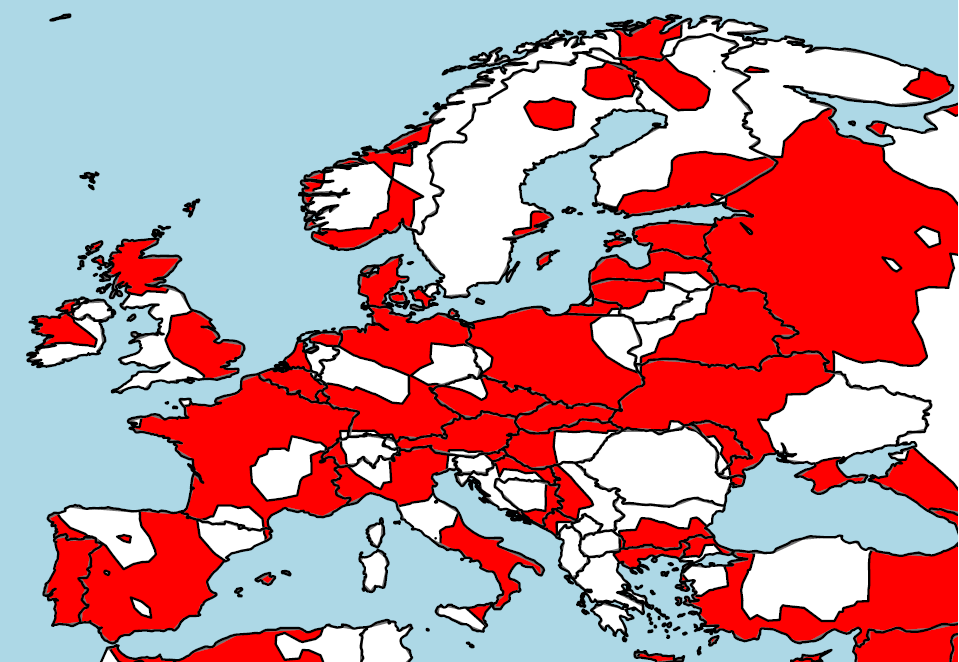}}
  \subfigure[$\alpha = 0.05$]{\includegraphics[height=4.8cm, width=4cm]{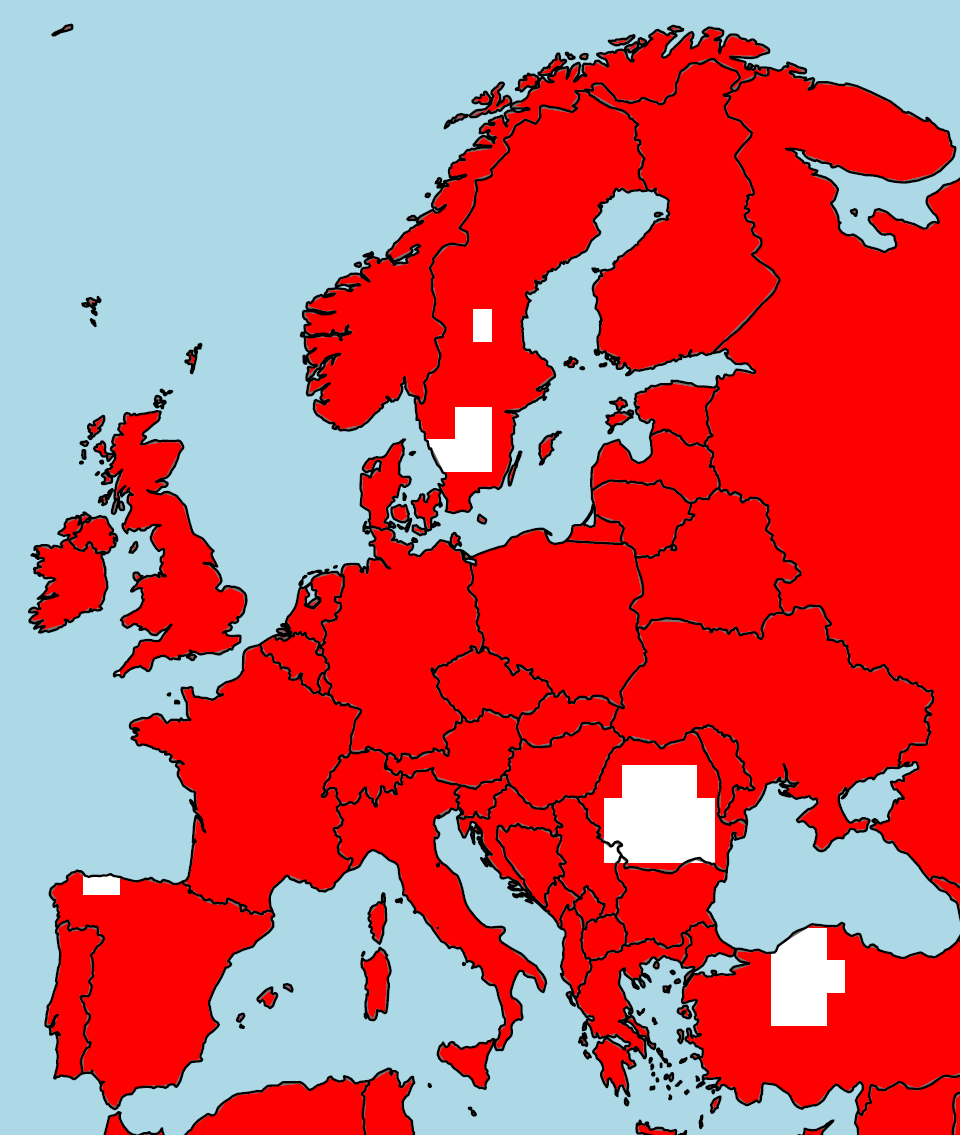}}
  \subfigure[$\alpha = 0.01$]{\includegraphics[height=4.8cm, width=4cm]{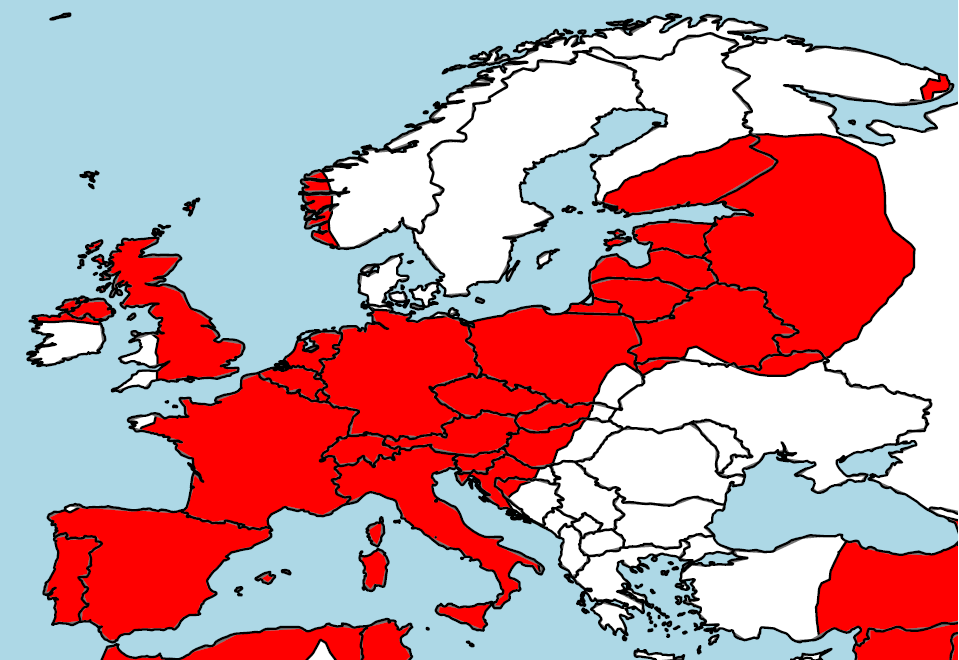}}
  \subfigure[$\alpha = 0.01$]{\includegraphics[height=4.8cm, width=4cm]{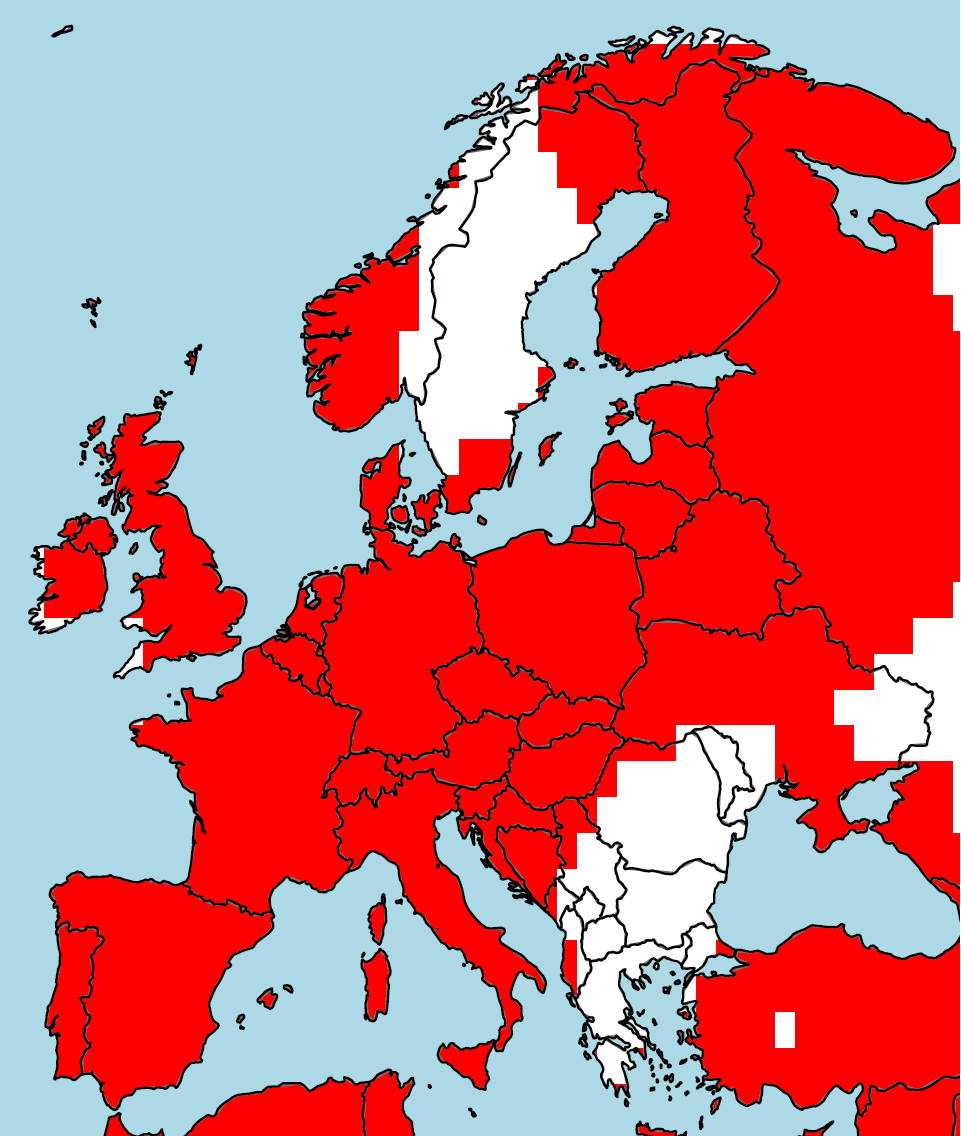}}
  \subfigure[$\alpha = 0.01$]{\includegraphics[height=4.8cm, width=4cm]{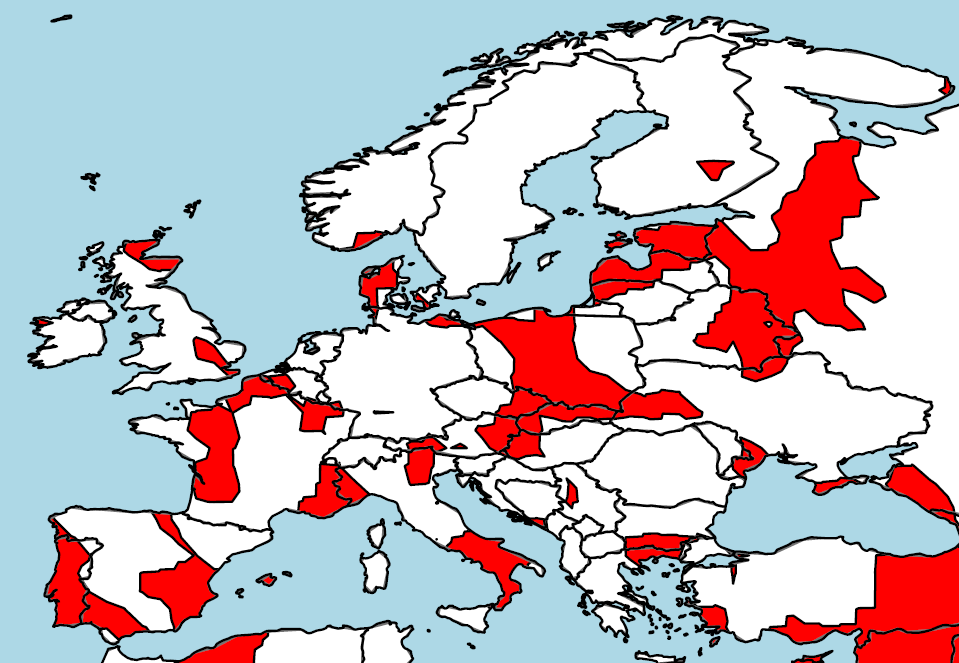}}
  \subfigure[$\alpha = 0.01$]{\includegraphics[height=4.8cm, width=4cm]{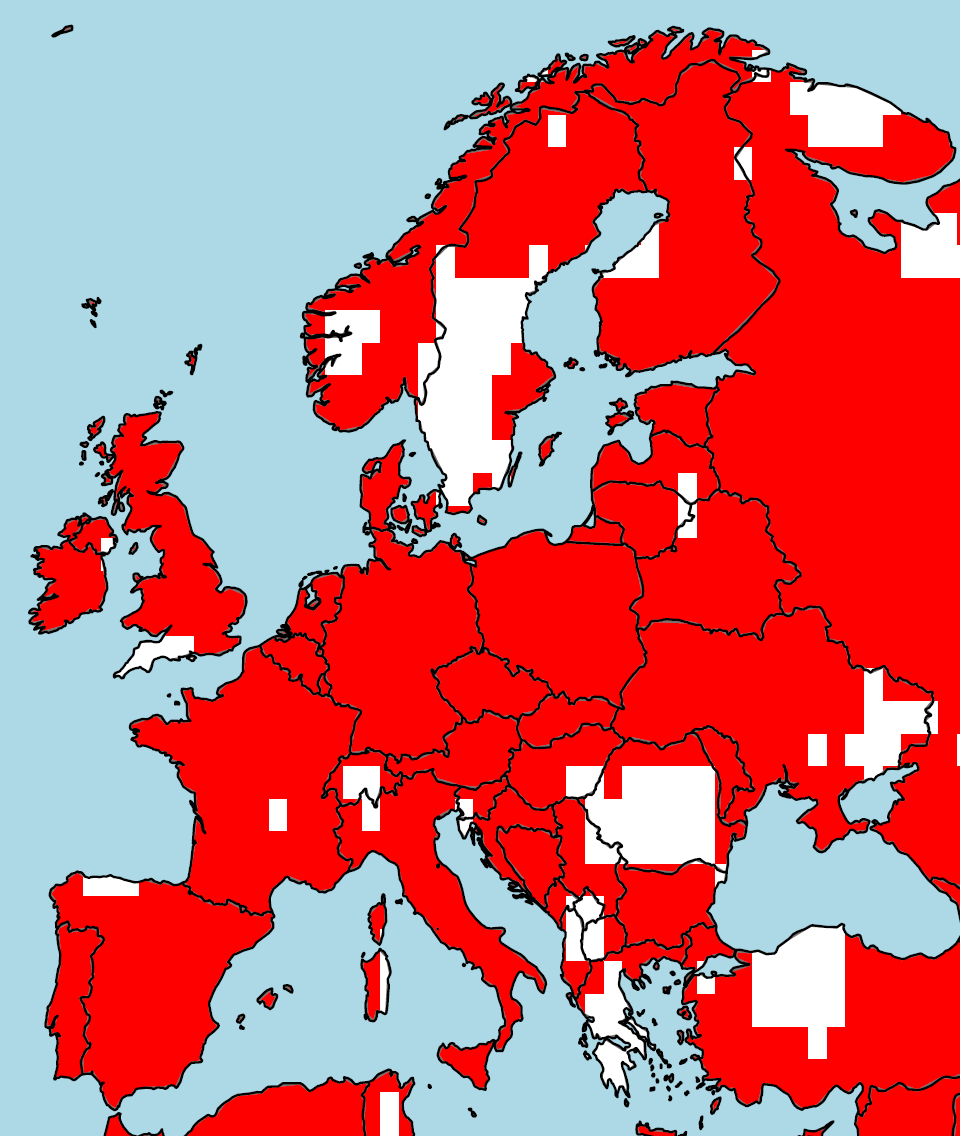}}
        \caption{Significance assessment of trend estimates. For comparison, all estimates have been extrapolated to the same 1 degree lattice-based grid prior to the significance assessment. The upper panels display results at the 5\% significance level, whereas 1\% significances are found in the bottom row. Avoidance excursion sets for simultaneous credible regions of no trend for a 5 degree data grid is shown in (a) and (e) while results based on a 1 degree data grid are shown in (c) and (g). Here, the avoidance set where the null hypothesis of no trend is rejected is indicated in red. Results based on marginal credible intervals for a 5 degree data grid are shown in (b) and (f), and for a 1 degree data grid in (d) and (h). }\label{fig:significance}
\end{figure}

In Figure~\ref{fig:significance} we compare a classical significance assessment of the trend estimates where a test is performed based on the marginal posterior distribution in each lattice cell independently and assessment based on excursion sets as proposed by \cite{BolinLindgren2015}. In the spatial test, the interpretation is that for a large set of realized trend fields, only 5\% of the fields should exhibit no trend anywhere within the avoidance set for $\alpha = 0.05$ and 1\% of the fields for $\alpha = 0.01$. This is a much stricter condition than that of assessing the marginal credible intervals. Consequently, the avoidance excursion sets are significantly smaller than the corresponding collection of grid cells where a null hypothesis of no trend is rejected. For example, the avoidance excursion set for a 5 degree data grid and $\alpha=0.05$ is similar in size and shape as the set for the marginal assessment of the same data at $\alpha = 0.01$. 

The avoidance excursion sets are widely different for the two data resolutions. At the higher data resolution, the avoidance excursion set has a smaller total area and a more irregular structure. However, the avoidance excursion set for the 1 degree data at a given $\alpha$ level is not a subset of the avoidance excursion set of the coarser data at the same $\alpha$ level, see for instance the area that comprises southern Bulgaria, eastern part of Greece and Turkey north of the Sea of Marmara.

For Fennoscandia and the Iberian Peninsula, we see a systematic decline in the size of the avoidance excursion set as the data resolution gets finer. For the 5 degree data, the entire Iberian Peninsula is included in the avoidance excursion set for the level $\alpha=0.05$, cf. Figure~\ref{fig:significance}(a), with less than half covered for the finest data resolution of 0.5 degree, cf. Figure~\ref{fig:Iberia}(c). While the data from Fennoscandia is generally found to exhibit less of a trend, a similar effect of the data resolution is apparent. However, note that for the coarsest data resolution of 5 degree and $\alpha=0.05$, the area around Stockholm on the east coast of Sweden is not included in the avoidance excursion set, cf. Figure~\ref{fig:significance}(a), while this small region comprises the entire avoidance excursion set at the finest data resolution of 0.5 degree, cf. Figure~\ref{fig:Fenno}(c). Considering the intermediate data resolution of 1 degree reveals that the Stockholm area is part of the avoidance excursion set at this stage as well, see Figure~\ref{fig:significance}(c). Furthermore, there is an area in the southern part of the Lappland district up north in the country that goes into the avoidance excursion set of the 1 degree resolution data, see again Figure~\ref{fig:significance}(c). Interestingly, this area is not included in the avoidance excursion set of neither the coarser 5 degree nor the fine-scale 0.25 degree resolution maps, cf. Figures~\ref{fig:significance}(a) and (c) and Figure~\ref{fig:Fenno}(c).

\subsection{Posterior distributions of hyperparameters}

A further assessment of the difference between the four analyses can be made through a comparison of the respective posterior distributions of the hyperparameters of the statistical model. The model has five hyperparameters, the first-lag autocorrelation coefficient and two parameters to describe the spatial structure of each of the two latent GRFs. For interpretability, we consider the posterior distributions of the marginal variance \eqref{eq:variance} and the range parameter \eqref{eq:range} for each GRF rather than those for the model parameters $\tau$ and $\kappa$. 

\begin{figure}[!hbpt]
  \centering
    \includegraphics[width=0.8\textwidth]{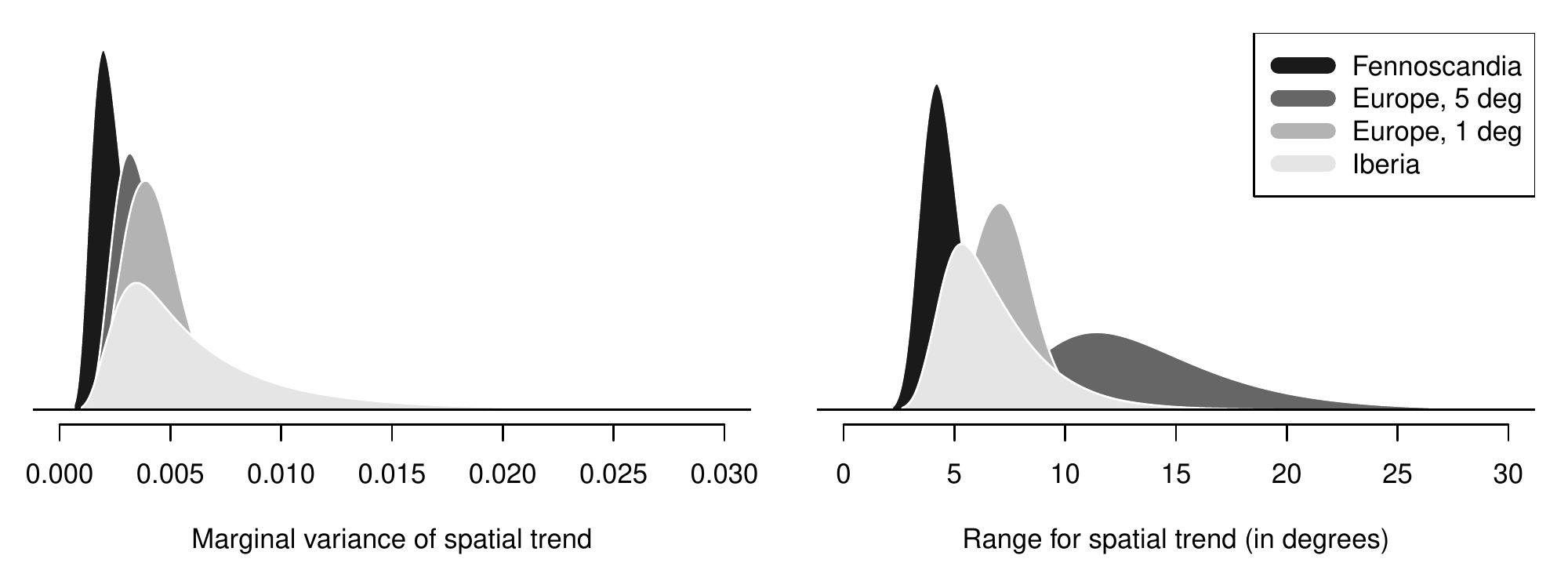}
  \caption{Posterior distributions of the marginal variance \eqref{eq:variance} and the range parameter \eqref{eq:range} for
  the spatial trend coefficient for all four data sets.}\label{fig:Parameters:trend}
\end{figure}

The posterior distributions for the marginal variance and the range parameter for the spatial trend coefficient are given in Figure~\ref{fig:Parameters:trend}. Recall that the range parameter is defined as the distance at which the correlation has fallen to approximately 0.13. The posterior mean estimates of the range are 13.4 degrees for the 5 degree data, 7.1 for the 1 degree data, 6.7 for the 0.5 degree data in Iberia and 4.4 for the 0.5 degree data in Fennoscandia. This indicates a substantial correlation between the trend estimates across neighboring grid cells irrespective of the data grid resolution, with the neighborhoods growing in size with finer data resolution. The posterior distributions for the marginal variances are somewhat more similar with the posterior mean of the marginal spread varying between 0.05 and 0.08. The uncertainty in the estimates of the marginal variance is much higher in Iberia than in Fennoscandia for identical data resolutions. While this might be an indication of structural non-stationarities across Europe, it is also worth noting that the Fennoscandia data set is roughly three times larger than that for Iberia, cf. Table~1. 

\begin{figure}[!hbpt]
  \centering
    \includegraphics[width=0.8\textwidth]{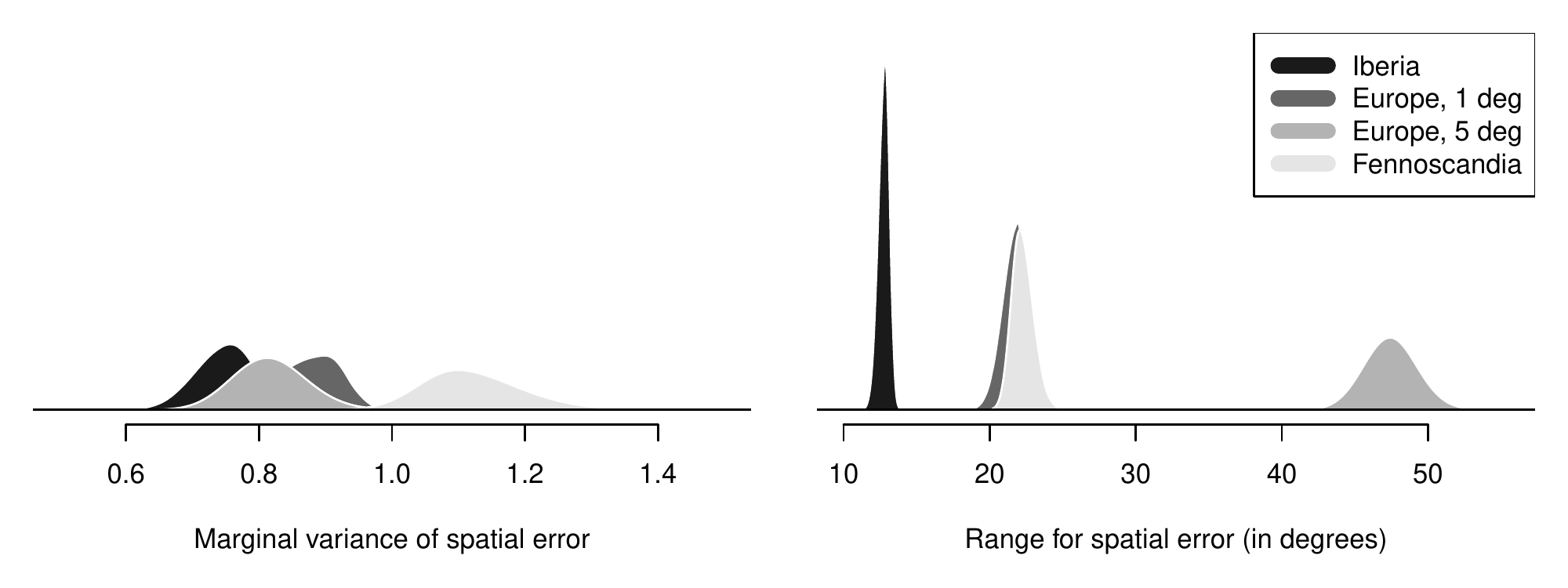}
  \caption{Posterior distributions of the marginal variance \eqref{eq:variance} and the range parameter \eqref{eq:range} for
  the spatially structured error term for all four data sets.}\label{fig:Parameters:error}
\end{figure}

The results for the spatial error terms are presented in Figure~\ref{fig:Parameters:error}. The spatial error field has a much larger range parameter than the spatial trend field, cf. Figure~\ref{fig:Parameters:trend}.  In particular, the range reaches across nearly the entire study region for the coarsest data resolution.  Here, the posterior mean of the marginal spread ranges from 0.86 for the Iberia data to 1.06 for the Fennoscandia data, and the posterior uncertainties are comparable. Note that the marginal variance estimates for the two latent fields cannot be compared as the associated units are different.   

\begin{figure}[!hbpt]
  \centering
    \includegraphics[width=0.6\textwidth]{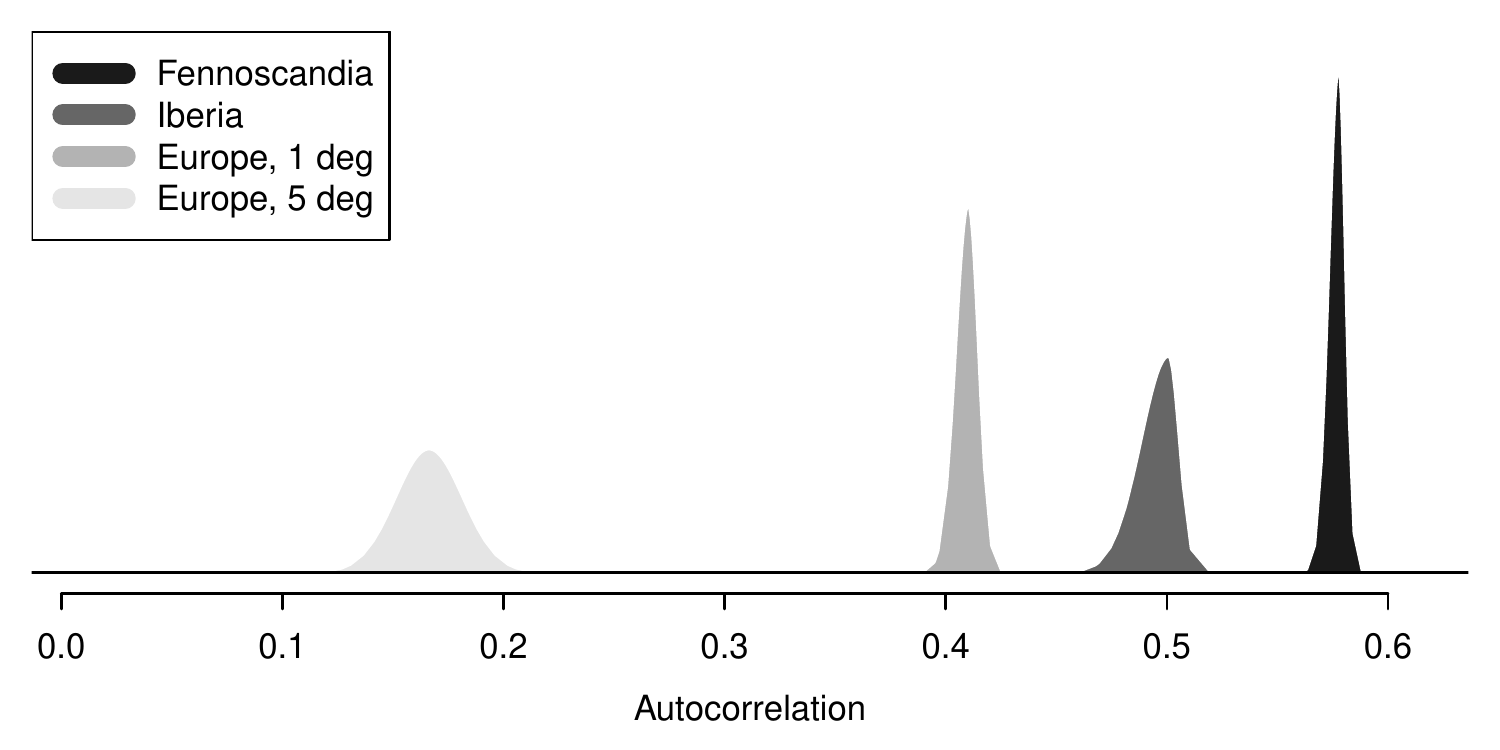}
  \caption{Posterior distributions of the first-lag autocorrelation coefficient for all four data sets.}\label{fig:Parameters:acf}
\end{figure}

The posterior distributions of the first-lag autocorrelation coefficient $\varphi$ are given in Figure~\ref{fig:Parameters:acf}. As expected, the posterior mean of $\varphi$ decreases with increased spatial averaging for a coarser grid resolution. $\varphi$ has a posterior mean of 0.42 for the European grid at 1 degree resolution while it reduces to 0.17 for the 5 degree resolution. At the finest resolution of 0.5 degree, the posterior mean is 0.50 for Iberia and 0.59 for Fennoscandia. Furthermore, these two posterior distributions are non-overlapping indicating that the assumption of a constant autocorrelation coefficient across all of Europe might generally not hold. However, it is not apparent that there is a direct relationship between the size of the data grid and the spread of the posterior distribution, as the distribution for the largest data grid (1 degree grid over Europe) has a fairly small spread.    

\section{Conclusions}\label{sec:discussion}

In this article we propose a spatial trend analysis approach for gridded
temperature data, adding spatial components to the univariate trend model used
by the IPCC \citep{AR5WG1Ch2}. While the model and inference approaches are
not new, the joint significance assessment of \citet{BolinLindgren2015} has, to
the best of our knowledge, not been used in this context before.  The latent
trend coefficient field is found to have a range far beyond the grid resolution of the data, warranting the spatial structure of the model. The avoidance excursion sets for the spatial significance assessment are overall much smaller than correponding sets resulting from marginal assessments, supporting previous claims on the need to protect against overstatement and overinterpretation of multiple-testing results in this setting \citep{wilks2016stippling}.  Controlling the false discovery rate, as suggested by \citet{wilks2006field, wilks2016stippling}, has been shown to work well when the model estimation is performed independently in each grid point location in a frequentist manner. However, its Bayesian interpretation is not entirely straight-forward \citep{Storey2003}. Rather, we have investigated using the Bonferroni correction to counteract the problem of multiple marginal comparisons \citep{Bonferroni1936}. This correction--which has been criticised for being conservative--resulted in no rejections of the null hypothesis of no trend.  

While the avoidance excursion sets overall decrease in areas with a finer data resolution, the finer resolution sets are not subsets of those obtained using a coarser data resolution with a higher degree of spatial smoothing. This result stresses the necessity of using context-appropriate data to answer questions related to climate change adaptation decision making whenever such data is available. Simultaneously, it emphasizes the point that any climate change assessment should be accompanied by the appropriate uncertainty quantification.

The model we have applied in this study is fairly simple in structure. When models of this type are estimated independently in each grid point location, it is commonly found that the first-order autocorrelation coefficient $\varphi$ varies over space. Our results in Figure~\ref{fig:Parameters:acf} for Fennoscandia and Iberia support this claim. However, we found that it was not feasible to include a latent spatial field structure for $\varphi$ under our inference scheme. The model currently includes two latent GRFs and adding the third proved computationally challenging. Similarly, we found it challenging to analyze data sets much larger than those studied here under the current model. Further generalization of the modeling structure may also be appropriate. For instance, \citet{Gneiting&2007} show that a non-separable covariance structure provides a better fit for spatial wind speed data, indicating that similar results may hold for temperatures.

\section*{Acknowledgments}

This work was funded by the KLIMAFORSK program of the Research Council of Norway through project nr. 229754, ``Long-range memory in Earth's climate response and its implications for future global warming''.  We acknowledge the E-OBS data set from the EU-FP6 project ENSEMBLES (\url{http://ensembles-eu.metoffice.com}) and the data providers in the ECA\&D project (\url{http://www.ecad.eu}). We are grateful to David Bolin, Finn Lindgren, Elias T. Krainski, Fabian Bachl and Magne Aldrin for providing helpful comments on the modeling approaches and their software implementations.

\bibliographystyle{chicago}

\end{document}